\documentstyle[12pt]{article}
\begin{document}

\tolerance=5000

\def\be{\begin{equation}}
\def\ee{\end{equation}}
\def\bea{\begin{eqnarray}}
\def\eea{\end{eqnarray}}
\def\nn{\nonumber \\}
\def\e{{\rm e}}

\def\SEH{S_{\rm EH}}
\def\SGH{S_{\rm GH}}
\def\AdS5{{{\rm AdS}_5}}
\def\S4{{{\rm S}_4}}

\  \hfill
\begin{minipage}{3.5cm}
September 2002 \\
\end{minipage}

\vfill

\begin{center}
{\large\bf Newton potential in deSitter braneworld}

\vfill

{\sc Shin'ichi NOJIRI}\footnote{nojiri@cc.nda.ac.jp}
and {\sc Sergei D. ODINTSOV}$^{\spadesuit}
$\footnote{
odintsov@tspu.edu.ru}

\vfill

{\sl Department of Applied Physics \\
National Defence Academy,
Hashirimizu Yokosuka 239-8686, JAPAN}

\vfill

{\sl $\spadesuit$
Lab. for Fundamental Study,
Tomsk State Pedagogical University,
634041 Tomsk, RUSSIA}

\vfill

{\bf ABSTRACT}

\end{center}

Using graviton correlator on deSitter (dS) brane in 5d AdS or dS bulk 
we calculate the four-dimensional Newton potential. For flat brane in AdS bulk
the Randall-Sundrum result is recovered. For flat brane in dS bulk 
the sign of subleading ($1/r^3$) term in Newton potential is negative if
compare 
with AdS bulk. In accordance with dS/CFT correspondence this indicates 
that dual CFT should be non-unitary (for example, higher derivative
conformal theory).

\vfill


\newpage

\noindent
1. There are natural reasons to study deSitter braneworld 
where brane is deSitter (dS) space and bulk may be arbitrary 
up to some extent. Indeed, first of all, recent astrophysical 
data indicate that our universe is (or will be ) in dS phase. 
Second, recently conjectured dS/CFT correspondence \cite{dscft,hull}
shows that dS space may be used to describe the dual CFT living 
on the brane. Moreover, gravity trapping occurs on dS brane for 
both AdS or dS bulks (see \cite{SN,localization}) like in the 
original Randall-Sundrum braneworld \cite{RS} where brane is flat 
and bulk is 5d AdS space. 

In the study of gravity trapping, the Newton potential on 
the brane is relevant object to search for. Starting from 
the quantum corrected Newton potential \cite{duff} in four 
dimensions and using AdS/CFT table and braneworld relation 
between 4d gravitational constant and 5d quantities it was possible 
to prove \cite{DL} the complementarity between AdS/CFT 
correspondence and Randall-Sundrum braneworld. 
Definitely, this sheds some light to the structure of dual CFT.
(Such complementarity is seen in another way: reincarnation of 
trace anomaly driven inflation \cite{alexei} in the new braneworld 
inflation \cite{NOZ,HHR}).

In the present letter we calculate the Newton potential 
(up to sub-leading term) in dS or flat brane when bulk is 
5d AdS or dS space.
First of all, we check that when brane is flat (bulk is AdS), 
the Newton potential coincides with the one found in 
refs.\cite{RS,DLS,GKR}.
However, for curved dS brane the Newton potential at short 
distances shows the drastically different behaviour if 
compare with result for 4d gravity.
For flat brane in dS bulk the sub-leading ($1/r^3$) correction 
to the Newton potential becomes negative if compare with the 
case of AdS bulk. As the corresponding one-loop correction from 
usual 4d conformal matter is positive, in accordance with predictions 
of dS/CFT correspondence \cite{dscft,hull} it indicates that 
dual CFT should be non-unitary (like higher derivative CFT).

\ 

\noindent
2. We will start from the deSitter brane in AdS bulk space. 
The initial action looks as:
\bea
\label{actionAdS}
S &=& S_{\rm EH}+S_{\rm GH}+S_{1} ,\quad 
S_{\rm EH} = -{1\over 16\pi G} \int d^{5}x \sqrt{ g }  
\left( R + {12\over l^2}\right)\; , \nn
S_{\rm GH} &=& -{1\over 8\pi G} \int d^{4}x \sqrt{\gamma} 
\; K \; , \quad S_{1}=
{3\over 8\pi G l'}\int d^{4}x\sqrt{\gamma} \; .
\eea
Here  the coefficient of $S_1$ is chosen to be arbitrary 
for a while. The 5-dimensional Euclidean anti deSitter space
(hyperboloid) is natural solution of above theory. The 
corresponding metric is given by 
\bea
\label{metricH5}
ds^{2}_{{\rm S}_5}&=&l^2 \left( dy^{2} 
+ \gamma_{ij} dx^{i}dx^{j} \right) \; \nn
\gamma _{ij} &=& \sinh^{2}y\; \hat{\gamma}_{ij} \; ,
\eea
where $\hat{\gamma_{ij}}dx^{i}dx^{j}$ 
describes the metric of S$_{4}$ with unit radius. 
The equation determing the radius of the brane 
has the following form \cite{gcorr}:
\be
\label{L1}
\left.\partial_y A\right|_{\rm on\, the\, brane}= {l \over l'}\ .
\ee 
Here $A$ is given by
\be
\label{L2}
A=\ln \sinh y - \ln \cosh \sigma\ .
\ee
If there is a brane at $y=y_0$, 
defining the radius of S$_4$ as $R_b=l\sinh y_0$, one obtains 
the following equation from (\ref{L1}):
\be
\label{slbr2H}
{1 \over R_b}\sqrt{1 + {R_b^2 \over l^2}}= {1 \over l'} \ .
\ee
When $R_b\to +\infty$, the r.h.s. of (\ref{slbr2H}) 
goes to ${1 \over l}$ and when $R_b\to 0$, the r.h.s. goes to the 
positive infinity. Then if 
\be
\label{HH1}
l'<l\ ,
\ee
there is a unique non-trivial solution. When $l'\to l - 0$, the 
brane radius tends to infinity and spherical brane becomes flat one. 

In (\ref{slbr2H}), the r.h.s. comes from the brane tension and 
the l.h.s. from gravity. When $l'>0$, the brane tension makes 
the radius of the brane smaller. Then Eq.(\ref{slbr2H}) tells 
that the gravity  tends to make the radius larger. We should 
also note that  l.h.s. is the monotonically decreasing 
function of $R_b$. Then if the radius of the brane becomes 
smaller than $R_b$ given by (\ref{slbr2H}), the contribution 
from gravity becomes larger that that from the brane tension. 
Then the radius of the brane tends to become larger. On the 
other hand, if the radius becomes larger than $R_b$, the 
contribution from the tension becomes dominant and the radius 
tends to become smaller. Then one can find the solution in 
(\ref{slbr2H}) is stable.

We now rewrite the bulk metric which includes the
effect of the perturbations $h_{ij}(y,x)$ as follows:
\be
\label{met}
ds^{2}=l^2 \left( dy^{2} 
+ \left(\gamma_{ij} + 
h_{ij}(y,x) \right) dx^{i}dx^{j} \right) \ .
\ee
$h_{ij}$ is transverse and traceless with respect to the metric
on the spherical spatial sections:
\bea
\label{eq1}
\hat{\gamma}^{ij}h_{ij}(y,x)=\hat{\nabla}^{i}h_{ij}(y,x)=0,
\eea
with $\hat{\nabla}$ being the covariant derivative
defined by the metric $\hat{\gamma}_{ij}$. 
Then the linearized Einstein equation has 
the following form:
\bea
\label{eq3}
\nabla^{2}h_{\mu\nu}= {2\over l^2}h_{\mu\nu}\ .
\eea
We now use the standard tensor spherical harmonics 
$H_{ij}^{(p)}(x)$ \cite{higuchi} to expand the metric 
perturbations
\bea
\label{eq3b}
\hat{\gamma}^{ij}H^{(p)}_{ij}(x)=\hat{\nabla}^{i}H_{ij}^{(p)}(x)
=0\ ,
\eea  
They are tensor eigenfunctions of the Laplacian:
\bea
\label{lapl}
\hat{\nabla}^2 H_{ij}^{(p)} =\left(2-p(p+4)\right) H_{ij}^{(p)} \; ,
\eea
where $p=2,3,\cdots$. As usual in quantum mechanics, 
the eigenvalue of 
$-\nabla^2$ can be regarded as the square of the momentum. 
Especially when $p$ is large, we can regard $p$ as the 
absolute value of the momentum. 
Besides $p$, $H_{ij}^{(p)}$ has integer indeces $k$, $l$, and $m$, 
which take values in the ranges $0\leq k\leq p$, $0\leq l \leq k$, 
and $-k\leq m \leq k$. Then for a fixed $p$, the degree of the 
degeneracy is defined by
\be
\label{dgdg}
\sum_{k=0}^p \sum_{l=0}^k \left(2k + 1\right) = {(p+1)(p+2)
(2p+3) \over 6}\ ,
\ee
which reduces to $p^3 \over 3$ for large $p$. 
One can also normalize $H_{ij}^{(p)}$ as
\be
\label{dgdg3}\int \sqrt{\gamma} d^4\Omega_{{\rm S}_4}
H_{ij(klm)}^{(p)} H_{i'j'(k'l'm')}^{(p')} 
=\delta^{pp'}\delta_{ii'}\delta_{jj'}\delta_{kk'}\delta_{ll'}
\delta_{mm'}\ .
\ee
Here $\sqrt{\gamma} d^4\Omega_{{\rm S}_4}$ is the volume element 
of S$_4$ with unit radius. The condition of the completeness is 
given by 
\be
\label{cmp}
{1 \over \sqrt{\gamma}}\delta^{ii'}\delta^{jj'}
\delta\left(\Omega - \Omega'\right)
=\sum_{p,k,l,m} H^{(p)}_{ij(klm)}(\Omega)H^{(p)}_{i'j'(k'l'm')}
(\Omega') \ .
\ee
Here $\Omega$ ($\Omega'$) expresses the coordinates of a point 
on S$_4$.

The degeneracy corresponding to the measure of the continuous 
momentum $q$ in the limit of $R_b\to \infty$ is:
\be
\label{dgdg2}
\sum_p {(p+1)(p+2)(2p+3) \over 6}\cdots \to 
{V_4 R_b^4 \over \left(2\pi\right)^4} \int V_3 q^3 dq \cdots\ .
\ee
Here $V_4$ and $V_3$ are the volumes of S$_4$ and S$_3$ with 
unit radius:
\be
\label{V4}
V_4={8\pi^2 \over 3}\ ,\quad V_3=2\pi^2\ .
\ee
Then Eq.(\ref{dgdg2}) tells 
\be
\label{dgdg2b}
{p^3 \over 3}dp \sim {R_b^4 q^3 dq \over 3}\ .
\ee
that is
\be
\label{dgdg2c}
q \sim {p \over R_b}\ .
\ee

The metric perturbations can be written as a sum of 
separable perturbations of the form
\bea
\label{AdSpert}
h_{ij}(y,x)=f^{{\rm AdS}}_{p}(y)H_{ij}^{(p)}(x) \; .
\eea
Then as in \cite{HHR,HHR2,gcorr}, the graviton 
correlator may be calculated as follows:
\bea
\label{L4}
\left< h_{ij}(x)h_{i'j'}(x') \right> &=& \sum_{p=2}^{\infty}
W_{iji'j'}^{(p)} (x,x') {1\over 2 T_{p}(y_0)}\; , \nn
T_{p}(y_0) &=& {l^3 \over 2\pi G} {1\over 32 l^4}
\left( {{f^{\rm AdS}}'_{p}(y_0) \over f^{\rm AdS}_{p}(y_0) }
+ 4 \coth y_0 - {6l \over l'} \right) \; ,\nn
W_{iji'j'}^{(p)} (x,x') &\equiv& 
H^{(p)}_{ij}(x)H^{(p)}_{i'j'}(x') \ .
\eea
>From the linearized Einstein equation (\ref{eq3}), 
we find $f^{\rm AdS}_p(y)$ satisfies the following equation:
\be
\label{fp1}
f_p''(y) - \left[-4+\left\{ p(p+3)+2\right\}\sinh^{-2} y\right]
f_p(y)=0\ .
\ee
Taking $f^{\rm AdS}_p(y)=\sinh^{1 \over 2}y 
F_p\left(\cosh y\right)$, 
we obtain 
\be
\label{fp2}
0=\left(1-z^2\right)F_p''(z) - 2z F_p'(z) 
+ \left[ {3 \over2}\cdot{5 \over 2}
 - {\left(p + {3 \over 2}\right)^2 \over 1 - z^2}\right]F_p\ .
\ee
Here $z=\cosh y$. The solution is given by the 
associated Legendre functions:
\be
\label{fp3}
F_p(z)=P^{-\left(p + {3 \over 2}\right)}_{3 \over 2}(z)
=P^{-\left(p + {3 \over 2}\right)}_{-{5 \over 2}}(z)\ ,
\quad Q^{-\left(p + {3 \over 2}\right)}_{3 \over 2}(z)
=Q^{-\left(p + {3 \over 2}\right)}_{-{5 \over 2}}(z)\ .
\ee
We now consider the limit $l'\to l - 0$ and we 
assume $1\ll |p| \ll |z|$. Then Eq.(\ref{fp2}) is reduced 
into 
\be
\label{fp4}
0= -z^2F_p''(z) - 2z F_p'(z) 
+ \left[{3 \over2}\cdot{5 \over 2} 
+ {p^2 \over z^2}\right]F_p(z)\ .
\ee
The solution of the above equation is given in terms of the  
deformed Bessel functions 
\be
\label{fp4b}
F_p = z^{-{1 \over 2}} \left(a I_2\left({p \over z}\right) 
+ b K_2\left({p \over z}\right)\right)\ .
\ee
When $z$ is large (or $i{p \over z}$ is small), the 
deformed Bessel functions behave as 
\bea
\label{fp5b}
I_2\left({p \over z}\right)&\sim& \left( {p \over 2z} \right)^2 
\left[ {1 \over 2} + {1 \over 6}\left({p \over 2z}\right)^2 
+ {1 \over 48} \left({p \over 2z}\right)^4 + \cdots \right]\ ,\\
K_2\left({p \over z}\right)&\sim&{1 \over \pi}
\left({p \over 2z}\right)^{-2}\left[ 1 
 - \left({p \over 2z}\right)^2 - \left({p \over 2z}\right)^4
\left(\gamma + \ln \left({p \over 2z}\right)\right)
+ \cdots \right]\ .\nonumber
\eea
Here $\gamma$ is the Euler number: $\gamma=0.57721\cdots$. 
Note that $K_2\left({p \over z}\right)$ gives the 
dominant contribution when $z$ is large. With $a=0$ we 
find in the limit of $R_b\to \infty$
\be
\label{fp7AdS0}
T_{p}(y_0) = {l^3 \over 2\pi G} {1\over 32 l^4}
\left\{ -{pl \over R_b}{K_2'\left(pl/2R_b\right) 
\over K_2\left(pl/2R_b\right)}-2\right\}\ .
\ee
When the momentum $q$ is 
Wick-rotated $q\to iq$, the modified Bessel function $K_2$ is 
replaced by the (second class) Bessel function or Neumann 
function $N_2$:
\be
\label{Tpinf1}
T_{p}(y_0) = {l^3 \over 2\pi G} {1\over 32 l^4}
\left\{ -{pl \over R_b}{N_2'\left(pl/2R_b\right) 
\over N_2\left(pl/2R_b\right)}-2\right\}\ .
\ee
For large $ql={pl \over R_b}$, $N_2\left(pl/2R_b\right)$ in 
(\ref{Tpinf1}) behaves as
\be
\label{N2}
N_2\left(pl/2R_b\right)\sim \sqrt{4 R_b \over \pi p l}
\cos\left({p l \over 2R_b} + {\pi \over 4}\right)\ .
\ee
Then we find
\be
\label{Tpinf2}
T_{p}(y_0) \sim {l^3 \over 2\pi G} {1\over 32 l^4}
\left\{ {pl \over R_b}\tan\left({pl \over 2R_b} + 
{\pi \over 4} \right) - {3 \over 2} \right\}\ .
\ee
Then for large $p$, $T_{p}$ vanishes at 
\be
\label{zero}
{pl \over 2R_b}={ql \over 2}\sim \left(n-{1 \over 4}\right)\pi
\ee
for large integer $n$. There appear infinite number of massive modes 
\be
\label{AdSmass}
m^2 \sim \left(n-{1 \over 4}\right)^2{\pi^2 \over l^2}\ .
\ee

When $pl/R_b$ is small 
\bea
\label{fp7AdS}
T_{p}(y_0) &=& {l^3 \over 2\pi G} {1\over 32 l^4}
\left\{ - {p^2 l^2 \over 2R_b^2} - 
{p^4 l^4 \over 4 R_b^4} \left(\gamma + {3 \over 4}
+ \ln {pl \over 2R_b}\right) + \cdots \right\} \nn
&=& -{l \over 128\pi G} \left\{ q^2 
+ {q^4 l^2 \over 2} \left(\gamma + {3 \over 4}
+ \ln {ql \over 2}\right) + \cdots \right\} \ .
\eea
If we define a scalar propagator $\Delta$ as 
\bea
\label{Delta}
&& \Delta(q) = \left\{ q^2 
+ {q^4 l^2 \over 2} \left(\gamma + {3 \over 4}
+ \ln {ql \over 2}\right) + \cdots \right\}^{-1}
= \Delta_0(q) + \Delta_{\rm KK}(q) 
+ {\cal O}\left(q^2\right) \ ,\nn
&& \Delta_0(q)\equiv {1 \over q^2}\ ,\quad
\Delta_{\rm KK}(q)= - {l^2 \over 2} \left(\gamma + {3 \over 4}
+ \ln {ql \over 2}\right) \ .
\eea
$\Delta_0(q)$ corresponds to the contribution from 
the zero mode, which is trapped on the brane and 
$\Delta_{\rm KK}(q)$ to that from the massive Kaluza-Klein 
(KK) modes \cite{GKR,DL}. (Note that 4d graviton is massive 
in constant curvature space, for a recent discussion
of appearence of such mass see \cite{porrati}).
Then the Newton potential is given by
\bea
\label{fp8bAdS}
V\left(\left|{\bf r}\right|\right)&\sim& 
{m_1 m_2 G^{(4)} \over \left(2\pi\right)^2}
\int d^3 q \e^{i{\bf q}\cdot{\bf r}}\left(\Delta_0(q) 
+ {4 \over 3}\Delta_{\rm KK}(q)\right) \nn
&=&{m_1 m_2 G^{(4)}}\left[{1 \over r} + {2l^2 \over 3r^3}\right]\ .
\eea
The factor ${4 \over 3}$ in the first line of (\ref{fp8bAdS}) 
appears since the KK modes are massive. In (\ref{fp8bAdS}), 
the four dimensional gravitational coupling $G^{(4)}$ is defined 
by 
\be
\label{L8}
G^{(4)}={G \over 2l}\ ,
\ee

In \cite{DL,GT,MVV,DLS}, the $r^{-3}$ loop correction from 
the CFT to the Newton potential has been calculated as follows
\be
\label{DSL1}
V(r)={G^{(4)} m_1 m_2 \over r}\left(1 + {\alpha G^{(4)}
\over r^2}\right)\ .
\ee
The coefficient $\alpha$ is given by 
\be
\label{DSL2}
\alpha={1 \over 45\pi}\left(12n_1 + 3n_{1/2} + n_0\right)\ .
\ee
Here $n_1$, $n_{1/2}$, and $n_0$ are the the numbers of 
vectors, (Majorana) fermions, and (real) scalars of the CFT. 
For ${\cal N}=4$ $SU(N)$ or $U(N)$ super Yang-Mills theory, 
we have $\left(n_1, n_{1/2}, n_0\right)
=\left(N^2,4N^2,6N^2\right)$ then $\alpha = {2N^2 \over 3\pi}$. 
On the other hand, by using the AdS/CFT and  braneworld 
relation, we have $N^2=\pi l^2 / G^{(4)}$ and therefore
\be
\label{DSL3}
\alpha G^{(4)}={2 l^2 / 3}\ ,
\ee
which reproduces the result in the Randall-Sundrum 
model \cite{RS,MVV,DLS}:
\be
\label{DSL4}
V(r)={G^{(4)} m_1 m_2 \over r}\left(1 + {2l^2
\over 3r^2}\right)\ .
\ee
The above expression also coincides with Eq.(\ref{fp8bAdS}). 
Thus, within our method where explicit form of spherical graviton
correlator is used, we reproduced the Newton potential
of RS brane in AdS bulk.

Finally in this section, we consider the case that 
${R_b \over l}$ is finite but $p$ is large.
This is the case of spherical brane. 
Since when $p$ is large, 
the associated Legendre function 
$P^{-p-{1 \over 2}}_{-{5 \over 2}}\left(z\right)$ 
\be
\label{Ein8}
P^{-p-{1 \over 2}}_{{3 \over 2}}\left(z\right)
\rightarrow {\e^{\left(-p - {1 \over 2}\right)\pi i}
\over \Gamma\left({3 \over 2} + p\right)}
\left({1 + z \over 1 - z}\right)^{-{p + {1 \over 2} 
\over 2}}\ .
\ee
one finds 
\be
\label{tp6AdS}
T_{p}(y_0) \to {p \over 64\pi Gl} {\cosh y_0 + 1 \over 
\sinh y_0}\ .
\ee
This tells that the potential of the gravity behaves as 
${1 \over r^2}$ rather than ${1 \over r}$ at the short distance. 
At the very short distances, any manifold can be regarded to 
be flat. 
For the pure 4d theory, 
for any (curved or flat) manifold, 
the graviton correlator behaves as 
\be
\label{Cr1}
\left< h_{ij}(x)h_{i'j'}(x') \right> 
\sim \int {d^4p \over p^2}\e^{ip\cdot\left(x-x'\right)}
\sim {1 \over \left|x-x'\right|^2}\ ,
\ee
which tells that the gravitational potential shows the  
Coulomb-like behavior in 4 dimensions. On the other side, 
Eq.(\ref{tp6AdS}) gives
at the short distances 
\be
\label{Cr2}
\left< h_{ij}(x)h_{i'j'}(x') \right> 
\sim \int {d^4p \over p}\e^{ip\cdot\left(x-x'\right)}
\sim {1 \over \left|x-x'\right|^3}\ .
\ee
The behavior of $\left|x-x'\right|^{-3}$ is nothing but the 
5-dimensional one. Therefore the potential of the gravity 
behaves as ${1 \over r^2}$ rather than ${1 \over r}$. Then the 
gravity force in the induced gravity becomes stronger at short 
distances than that in the pure 4d case. 

\ 

\noindent
3. After the explicit calculation of Newton potential for 
flat and dS branes in AdS bulk in the previous section,
we will discuss such the brane in dS bulk space. 
We start with the following action:
\bea
\label{actionL}
S &=& S_{\rm EH}+S_{\rm GH}+S_{1} ,\quad 
S_{\rm EH} = -{1\over 16\pi G} \int d^{5}x \sqrt{ -g }  
\left( R -{12\over l^2}\right)\; , \nn
S_{\rm GH} &=& -{1\over 8\pi G} \int d^{4}x \sqrt{\gamma} 
\;\; , \quad S_{1}= -
{3\over 8\pi G l'}\int d^{4}x\sqrt{\gamma} \ .
\eea
Here the coefficient of $S_1$ is chosen to be arbitrary 
for a while.
There exists the 5-dimensional Lorentzian deSitter space solution:  
\bea
\label{metricS5L}
ds^{2}_{{\rm dS}_5}&=&l^2 \left( - dy^{2} 
+ \gamma_{ij} dx^{i}dx^{j} \right) \; \nn
\gamma _{ij} &=& \cosh^{2}y\; \hat{\gamma}_{ij} \; ,
\eea
which can be obtained from the following metric of 
5-dimensional sphere $S^{5}$:  
\bea
\label{metricS5}
ds^{2}_{{\rm S}_5}&=&l^2 \left( dy^{2} 
+ \gamma_{ij} dx^{i}dx^{j} \right) \; \nn
\gamma _{ij} &=& \sin^{2}y\; \hat{\gamma}_{ij} \; ,
\eea
by analytically continuing $y$ as
\be
\label{an}
y\to {\pi \over 2} - iy\ .
\ee
Then from the equation determining the radius of the brane 
(\ref{L1})
\be
\label{L2dS}
A=\ln \cosh y - \ln \cosh \sigma\ ,
\ee
 defining the radius $R_b$ of the brane as $R_b=l\cosh y_0$, 
and assuming that there is a brane at $y=y_0$, we find
\be
\label{L3}
{1 \over R_b}\sqrt{{R_b^2 \over l^2}-1}={1 \over l'}\ .
\ee
Eq.(\ref{L3}) has a solution labeled by $l'$ if ${l \over l'}<1$ 
\cite{SN}:
\be
\label{L4b}
R_b={ll' \over \sqrt{{l'}^2 - l^2}}\ .
\ee
Then $R_b$ goes to infinity when ${l \over l'}\to 1 - 0$.

As in (\ref{slbr2H}), the r.h.s. in (\ref{L3}) expresses the 
contribution from the brane tension and the l.h.s. from gravity. 
The brane tension tends to make the radius of the brane smaller 
and the gravity tends to make the radius larger, again. 
We should note, however, in contrast with the case of 
(\ref{slbr2H}), that the l.h.s. is the monotonically increasing 
function of $R_b$ if $R_b>l$. Then if the radius of the brane 
becomes smaller than $R_b$ given by (\ref{L3}), the contribution 
from the tension becomes dominant and the radius tends to become 
further smaller. On the other hand, if the radius becomes larger 
than $R_b$, the contribution from gravity becomes larger and the 
radius tends to become further larger. Therefore,  the solution 
in (\ref{L3}) is instable in contract with that of 
(\ref{slbr2H}).  This indicates also that dual CFT should be non-unitary
theory.

We now rewrite bulk metric which includes the
effect of the perturbations $h_{ij}(y,x)$ as in (\ref{met}):
\be
\label{metdS}
ds^{2}=l^2 \left( - dy^{2} 
+ \left(\gamma_{ij} + 
h_{ij}(y,x) \right) dx^{i}dx^{j} \right) \ .
\ee
(The relevance of metric perturbations in dS/CFT correspondence 
has been recently discussed in \cite{abdalla}).
Then the linearized Einstein equation has the following form:
\bea
\label{eq3dS}
\nabla^{2}h_{\mu\nu}=-{2\over l^2}h_{\mu\nu}\ .
\eea
As in (\ref{AdSpert}), the metric perturbations can be written 
as a sum of separable perturbations of the form
\bea
\label{dSpert}
h_{ij}(y,x)=f_{p}(y)H_{ij}^{(p)}(x) \; .
\eea
The graviton 
correlator may be calculated as follows \cite{gcorr}:
\bea
\label{L4dS}
\left< h_{ij}(x)h_{i'j'}(x') \right> &=& \sum_{p=2}^{\infty}
W_{iji'j'}^{(p)} (x,x') {1\over 2 T_{p}(y_0)}\; , \nn
T_{p}(y_0) &=& -{l^3 \over 2\pi G} {1\over 32 l^4}
\left( {f'_{p}(y_0) \over f_{p}(y_0) }+ 4 \tanh y_0
 - {6l \over l'} \right) \; .
\eea
>From the linearized Einstein equation (\ref{eq3dS}), 
 $f_p(y)$ satisfies the following equation:
\be
\label{fp1dS}
f_p''(y) + \left[-4+\left\{ p(p+3)+2\right\}\cosh^{-2} y\right]
f_p(y)=0\ .
\ee
With $f_p(y)=\cosh^{1 \over 2}y F_p\left(i\sinh y\right)$, 
we obtain (\ref{fp2}) again with $z=i\sinh y$. Then the solution 
is again given in terms of the associated Legendre functions. 
In the limit ${l \over l'}\to 1 - 0$,  the solution 
is given by the  Bessel functions 
\be
\label{fp4bdS}
F_p = z^{-{1 \over 2}} \left(a J_2\left(i{p \over z}\right) 
+ b N_2\left(i{p \over z}\right)\right)\ .
\ee
When $z$ is large (or $i{p \over z}$ is small), the Bessel 
functions behave as 
\bea
\label{fp5bdS}
J_2\left(i{p \over z}\right)&\sim& \left( i{p \over 2z} \right)^2 
\left[ {1 \over 2} - {1 \over 6}\left(i{p \over 2z}\right)^2 
+ {1 \over 48} \left(i{p \over 2z}\right)^4 + \cdots \right]\ ,\\
N_2\left(i{p \over z}\right)&\sim&{1 \over \pi}
\left(i{p \over 2z}\right)^{-2}\left[ 1 
+ \left(i{p \over 2z}\right)^2 - \left(i{p \over 2z}\right)^4
\left(\gamma + \ln \left(i{p \over 2z}\right)\right)
+ \cdots \right]\ .\nonumber
\eea
when $z$ is large, $N_2\left(i{p \over z}\right)$ gives the 
dominant contribution. If we put $a=0$, we 
find in the limit of $R_b\to \infty$
\be
\label{Tpinf1dS}
T_{p}(y_0) = {l^3 \over 2\pi G} {1\over 32 l^4}
\left\{ -{pl \over R_b}{N_2'\left(pl/2R_b\right) 
\over N_2\left(pl/2R_b\right)}-2\right\}\ .
\ee
Then when $pl/R_b$ is small 
\bea
\label{fp7dS}
T_{p}(y_0) &=& {l^3 \over 2\pi G} {1\over 32 l^4}
\left\{ {p^2 l^2 \over 2R_b^2} - 
{p^4 l^4 \over 4 R_b^4} \left(\gamma + {3 \over 4} 
+ \ln {pl \over 2R_b}\right) + \cdots \right\} \nn
&=& {l \over 126\pi G} \left\{ q^2 - 
{q^4 l^2 \over 2} \left(\gamma + {3 \over 4} 
+ \ln {ql \over 2}\right) + \cdots \right\}\ .
\eea
Using similar way as in previous section, 
the Newton potential is found to be
\be
\label{fp8}
V\left(\left|{\bf r}\right|\right)\sim
m_1 m_2 G^{(4)}\left[{1 \over r} - {2l^2 \over 3r^3}\right]\ .
\ee
We should note that the sign of the $1/r^3$ correction term 
is different from that in (\ref{fp8bAdS}). 

For large $ql={pl \over R_b}$, $N_2\left(pl/2R_b\right)$ in 
(\ref{Tpinf1dS}) behaves as in (\ref{N2}). Then we find
\be
\label{Tpinf2dS}
T_{p}(y_0) \sim {l^3 \over 2\pi G} {1\over 32 l^4}
\left\{ {pl \over R_b}\tan\left({pl \over 2R_b} + 
{\pi \over 4} \right) - {3 \over 2} \right\}\ .
\ee
For large $p$, $T_{p}$ vanishes at 
\be
\label{zerodS}
{pl \over 2R_b}={ql \over 2}\sim \left(n-{1 \over 4}\right)\pi
\ee
for large integer $n$. 
Since $q$ is spacelike or Euclidean momentum, Eq.(\ref{zerodS}) 
tells that there appear the tachyon modes with masses, 
which corresponds to (\ref{AdSmass}),  
\be
\label{tmass}
m^2 \sim - \left(n-{1 \over 4}\right)^2{\pi^2 \over l^2}\ .
\ee
This is consistent with the instability of  solution 
in (\ref{L3}). 
Then the field theory on the brane seems to be non-unitary. 
Indeed, Strominger \cite{dscft} and Hull \cite{hull} have conjectured that
CFT in dS/CFT 
correspondence is not unitary. Then the above result seems 
to be consistent. Since the usual field theory gives 
positive contribution to the $1/r^3$ corrections to 
the Newton potential \cite{duff}, then the dual field theory 
on the brane might be higher derivative conformal theory 
like higher derivative scalar, which is not unitary in general. 
Indeed,  for the higher derivative conformal
scalar  $\alpha=-{8 \over 45\pi}$.

In the above argument, the limit $1\ll |p| \ll |z|={R_b \over l}$ 
is considered. Let us consider the case that 
${R_b \over l}$ is finite but $p$ is large. 
Since when $p$ is large, we find 
\be
\label{tp6}
T_{p}(y_0) \to {p \over 64\pi Gl} \tanh y_0 \ ,
\ee
the potential of the gravity behaves as 
${1 \over r^2}$ rather than ${1 \over r}$ at  short distance 
as in (\ref{Cr2}). The leading behavior is not changed with 
that in the case that the bulk is AdS although the sub-leading 
behavior might be changed. It is not easy to compare such 
subleading behavior with the experimental data, when 
the energy is very large. Furthermore in general there 
appears the anomaly induced effective action on the brane 
\cite{NOZ,HHR}, which changes the short distance behavior 
drastically \cite{HHR,HHR2,gcorr}. Then it would be 
difficult to predict if the bulk is dS or AdS from 
the structure of Newton potential.

\ 

\noindent
{\bf Acknowledgments}

We thank J. Isern for interest in this work.
The work by S.N. is supported in part by the Ministry of Education, 
Science, Sports and Culture of Japan under the grant n. 13135208.

\end{document}